%% file: main.tex
\documentclass[a4paper,11pt]{article}
\usepackage{jinstpub}
\usepackage{lineno}

\usepackage{booktabs}
\usepackage{xcolor}
\usepackage{tikz}
\usetikzlibrary{arrows.meta,positioning,fit}

\newcommand{\todo}[2][]{%
  \textcolor{red}{\textbf{TODO\ifx\relax#1\relax\else[#1]\fi:}\ #2}}

\title{\boldmath NuGlass: A Fast and Interactive 3D Visualization of Neutrino
Oscillations}

\author{Chao Zhang\note{Corresponding author.}}
\affiliation{Brookhaven National Laboratory,\\Upton, NY 11973, USA}

\emailAdd{czhang@bnl.gov}

\abstract{Neutrino oscillation phenomenology is usually
communicated through two-dimensional projections (probability curves,
oscillograms, and bi-probability plots), each of which flattens a
function of many parameters (baseline, energy, matter density, mass
ordering, and the CP-violating phase $\delta_{\rm CP}$) onto a single
static figure. We present NuGlass, an open-source, web-based
application that renders six complementary, fully interactive 3D views of
the same underlying three-flavor evolution: an oscillogram surface, a
bi-probability trajectory extruded along energy, a Bloch-sphere projection of the quantum flavor state, a matter-eigenstate
phasor decomposition of the oscillation amplitude, a flavor-composition
tube along the propagation baseline, and a worldline view that plots the
probabilities against $L/E$ and makes the matter-induced breaking of the
vacuum $L/E$ degeneracy visible. A
global channel switcher ($\nu_\mu\to\nu_e$, $\nu_\mu\to\nu_\mu$,
$\nu_\mu\to\nu_\tau$, $\nu_e\to\nu_e$), experiment presets, and customizable animation controls are provided.
NuGlass can be used
both as a research exploration tool and as a public-engagement
exhibit.
The application runs entirely client-side and is deployed at \url{https://czczc.github.io/nuglass/}.}


\begin{document}
\maketitle
\flushbottom

\input{sections/01-introduction.tex}
\input{sections/02-physics-engines.tex}
\input{sections/03-views.tex}
\input{sections/04-architecture.tex}
\input{sections/05-insights.tex}
\input{sections/06-discussion.tex}
\input{sections/07-conclusions.tex}

\acknowledgments

The web application was developed with the assistance of Anthropic's
Claude Code (Claude Fable 5 model) through the workflow described in
section~\ref{sec:workflow}. AI assistance was used for language
refinement of this manuscript. This work is supported by the US Department of Energy (DOE) Office of Science, Office of High
Energy Physics under Contract No. DE-SC0012704.

\bibliographystyle{JHEP}
\bibliography{biblio}

\end{document}

%% file: sections/01-introduction.tex
\section{Introduction}
\label{sec:intro}

Neutrino oscillation is intrinsically a high-dimensional phenomenon. The oscillation probability depends simultaneously on the
baseline $L$, the neutrino energy $E$, the matter density $\rho$ along the
path, the neutrino/antineutrino channel, the mass ordering, and six
oscillation parameters. Among these, the CP-violating phase
$\delta_{\rm CP}$ and the mass ordering are the primary measurement
targets of the coming decade~\cite{dune-tdr,nova,t2k,juno}.

This phenomenology is often communicated through
static two-dimensional projections. Probability-versus-energy curves show
one baseline at a time; $L/E$ oscillograms fix $\delta_{\rm CP}$ and the
ordering; bi-probability diagrams~\cite{minakata-nunokawa}
display the $(\nu, \bar\nu)$ correlation but compress the energy
dependence into a one-parameter family of ellipses. Each projection is
well adapted to the question it was designed for, but hides the
dependencies it marginalizes over. For students entering the field,
non-expert collaborators, and the general public, it is therefore
difficult to assemble a mental model of the full parameter dependence
from a collection of 2D figures. Even for experienced researchers, some structural
features of the three-flavor evolution (discussed in
section~\ref{sec:insights}) are difficult to see in standard
projections.

Modern web browsers make it practical to go beyond static 2D figures.
WebGL provides hardware-accelerated 3D rendering on essentially every
laptop and phone, reactive UI frameworks make it straightforward to bind
physics parameters to live redraws, and the numerical problem itself is
well within JavaScript's performance envelope (discussed in section~\ref{sec:engines}).
This paper presents \emph{NuGlass}, an open-source web application that
provides six linked, fully interactive 3D views of three-flavor
oscillations in constant-density matter, backed by two independently
implemented and mutually cross-validated physics engines. The name
combines $\nu$ (``nu''), the symbol for the neutrino, with ``glass'' in
the sense of a looking glass: the application is intended as a lens
through which the multi-dimensional structure of neutrino oscillations
can be examined from any angle. The application
targets researchers with exact engines, global-fit defaults, experiment
presets, and quantitative hover readouts; students with linked views of
the same quantum state and amplitude-level decompositions; and the
general public with animation-first defaults, outreach-level
documentation, and a web deployment that requires no installation.

Many numerical libraries exist to calculate the neutrino oscillation
probabilities, such as GLoBES~\cite{globes}, Prob3++~\cite{prob3pp},
OscProb~\cite{oscprob}, NuOscProbExact~\cite{nuoscprobexact},
nuSQuIDS~\cite{nusquids}, nuosc~\cite{nuosc}, and the NuFast
reference implementation~\cite{nufast}. On the visualization side,
browser-based animations of the flavor-fraction evolution exist as
outreach projects (e.g., VISOS~\cite{visos}), and interactive
Bloch-sphere simulators are common in the quantum-information community,
but we are not aware of one applied to neutrino flavor evolution in
matter. To our knowledge, NuGlass is the first tool
to combine (a)~exact amplitude-level computation in the web browser,
(b)~multiple linked 3D geometric representations, and (c)~experiment presets with user-friendly customization tools.

This paper is organized as follows. Section~\ref{sec:engines} describes
the underlying physics formalism, the two computation engines, and their
cross-validation. Section~\ref{sec:views} presents the six 3D views.
Section~\ref{sec:architecture} summarizes the application architecture
and performance. Section~\ref{sec:insights} develops two example physics
observations enabled by the 3D representations.
Section~\ref{sec:discussion} discusses the roles and limitations of the
application, and section~\ref{sec:conclusions} concludes.

%% file: sections/02-physics-engines.tex
\section{Physics engines and parameter defaults}
\label{sec:engines}

\subsection{Formalism and engines}
\label{sec:formalism}

NuGlass computes three-flavor oscillations in matter of constant density.
The flavor-basis Hamiltonian is
\begin{equation}
  H = \frac{1}{2E}\, U\, \mathrm{diag}(0,\ \Delta m^2_{21},\ \Delta m^2_{31})\, U^\dagger
      + \mathrm{diag}(\pm V,\ 0,\ 0),
  \label{eq:hamiltonian}
\end{equation}
with $U$ the PMNS matrix parametrized by $\theta_{12}$, $\theta_{13}$,
$\theta_{23}$, and $\delta_{\rm CP}$~\cite{pmns-pontecorvo,pmns-mns}, and
$V = \sqrt{2}\, G_F N_e$ the Wolfenstein matter
potential~\cite{wolfenstein,msw} with electron density
$N_e \propto Y_e\, \rho$. An initial flavor state $|\nu_\alpha\rangle$
evolves over a baseline $L$ as
$|\nu(L)\rangle = e^{-iHL}\,|\nu_\alpha\rangle$. Diagonalizing
eq.~(\ref{eq:hamiltonian}) as
$H = W\,\mathrm{diag}(\lambda_1,\lambda_2,\lambda_3)\,W^\dagger$, with
$W$ the mixing matrix and $\lambda_i$ the eigenvalues in matter
($W \to U$ and $\lambda_i \to \Delta m^2_{i1}/2E$ in vacuum), the two
quantities NuGlass visualizes are the complex flavor amplitudes and the
oscillation probabilities,
\begin{equation}
  a_\beta(L) = \langle\nu_\beta|\, e^{-iHL}\, |\nu_\alpha\rangle
             = \sum_i W_{\beta i}\, W^*_{\alpha i}\, e^{-i\lambda_i L},
  \qquad
  P(\nu_\alpha\to\nu_\beta) = \left|a_\beta(L)\right|^2.
  \label{eq:amplitude}
\end{equation}
Every quantitative construction in section~\ref{sec:views} derives from
eq.~(\ref{eq:amplitude}): the Bloch vector of eq.~(\ref{eq:bloch}) is
built from the $a_\beta$, the phasor sum of eq.~(\ref{eq:phasors}) is
its term-by-term decomposition, and the vacuum probability of
eq.~(\ref{eq:vacuum}) is $|a_\beta|^2$ evaluated at $W \to U$.

In practical units, the kinematic phases and the matter term
($a$, the potential of eq.~(\ref{eq:hamiltonian}) expressed in the same
${\rm eV}^2$ units as the mass splittings) are
\begin{align}
  \Delta_{ij} &\equiv \frac{\Delta m^2_{ij} L}{4E}
    = 1.267\, \frac{\Delta m^2_{ij}}{{\rm eV}^2}\,
      \frac{L}{\rm km}\, \frac{\rm GeV}{E},
  \label{eq:phase}\\
  a &\equiv 2EV = 2\sqrt{2}\, G_F N_e E
    = 1.52588\times10^{-4}~{\rm eV}^2 \cdot
      \frac{Y_e\, \rho}{\rm g\,cm^{-3}}\, \frac{E}{\rm GeV},
  \label{eq:matterterm}
\end{align}
so vacuum phases depend on $L$ and $E$ only through $L/E$, while the
matter term grows with $E$ alone. Antineutrinos are obtained by
$\delta_{\rm CP} \to -\delta_{\rm CP}$, $V \to -V$; the inverted mass
ordering is approximated by flipping the sign of $\Delta m^2_{31}$ at
fixed magnitude (section~\ref{sec:discussion} discusses this
approximation).


Two physics engines drive the underlying calculations. The primary probability engine is a line-by-line JavaScript port of
NuFast-LBL~\cite{nufast}, which computes all nine oscillation
probabilities in constant-density matter using real arithmetic only, with
the matter eigenvalues obtained from a single Newton refinement of an
analytic seed (configurable $N_{\rm Newton}$). The port preserves the
upstream sign conventions (antineutrinos as $E < 0$) and the MIT license
of the original.


Probability-only algorithms cannot supply the complex amplitudes needed
by the amplitude-level views. The second
engine therefore constructs the full complex Hamiltonian of
eq.~(\ref{eq:hamiltonian}) and diagonalizes it with a cyclic Jacobi
eigensolver for complex Hermitian $3\times3$ matrices. From the eigenpairs
$\{\lambda_i, W_i\}$ it provides:
\begin{itemize}
\item the complex flavor amplitudes $a_\beta(L)$ of
  eq.~(\ref{eq:amplitude}), used by the Statesphere and Flavortube
  views;
\item matter-eigenstate phasor terms
  $c_i = W_{\beta i} W^*_{\alpha i}$ with phase velocities $\lambda_i$,
  i.e.~the individual terms of eq.~(\ref{eq:amplitude}),
  used by the Phasors view (sections~\ref{sec:phasors}
  and~\ref{sec:phasor-insight});
\item probabilities $|a_\beta|^2$ as a by-product, used for
  cross-validation.
\end{itemize}


The two engines are compared on a dense grid: 25~energies
$\times$ 21~baselines $\times$ 5~densities $\times$ 4~CP phases
$\times$ $\nu/\bar\nu$ $\times$ both orderings (42\,000 parameter
points), with all nine flavor-channel probabilities compared at each
point (378\,000 comparisons). Table~\ref{tab:xval} lists the maximum
absolute probability difference as a function of the NuFast Newton
refinement count: $2.3\times10^{-7}$ at the app's default
$N_{\rm Newton} = 1$ and $2.3\times10^{-10}$ at $N_{\rm Newton} = 2$,
well below anything resolvable in the visualization.

\begin{table}[htbp]
  \centering
  \caption{Engine cross-validation. Maximum absolute difference between
  the NuFast-LBL port and the Jacobi amplitude engine over all nine
  oscillation channels on a grid of $E \in [0.2, 6]$~GeV (25 points),
  $L \in [0, 5000]$~km (21), $\rho \in \{0, 1.5, 2.85, 4.5, 8\}$~g/cm$^3$,
  $\delta_{\rm CP} \in \{0^\circ, 90^\circ, 212^\circ, 270^\circ\}$, both
  mass orderings, and both $\nu/\bar\nu$, at the NuFit~6.1 best-fit
  parameters.}
  \label{tab:xval}
  \smallskip
  \begin{tabular}{cc}
    \toprule
    $N_{\rm Newton}$ & $\max |\Delta P|$ \\
    \midrule
    0 & $6.6\times10^{-4}$ \\
    1 (app default) & $2.3\times10^{-7}$ \\
    2 & $2.3\times10^{-10}$ \\
    \bottomrule
  \end{tabular}
\end{table}

\subsection{Parameter defaults and experiment presets}
\label{sec:presets}

Oscillation parameters default to the NuFit~6.1 (2025) global-fit
best-fit point for normal ordering including IceCube-2024 and
Super-Kamiokande atmospheric data~\cite{nufit}: $\theta_{12} =
33.76^\circ$, $\theta_{13} = 8.62^\circ$, $\theta_{23} = 43.29^\circ$,
$\delta_{\rm CP} = 212^\circ$, $\Delta m^2_{21} = 7.537\times10^{-5}$~eV$^2$,
$\Delta m^2_{31} = +2.511\times10^{-3}$~eV$^2$. All six parameters are
user-adjustable through an ``all parameters'' panel.

A channel selector in the header sets the oscillation channel shown by every view:
$\nu_\mu\to\nu_e$, $\nu_\mu\to\nu_\mu$, $\nu_\mu\to\nu_\tau$, or
$\nu_e\to\nu_e$ (the initial flavor follows the channel). Six experiment
presets, summarized in table~\ref{tab:presets}, set the baseline and the source energy window~\cite{dune-tdr,nova,t2k,juno,kamland,dayabay}, and
declare which channels the experiment measures; the reactor presets also
switch the source to $\bar\nu$, and the axes and slider steps switch
to MeV at reactor scales. The parameter ranges in every view are scoped to the
active preset (e.g., $L$ sweeps span 0--2$\times$baseline), so each
experiment is always viewed on its own natural scales.

\begin{table}[htbp]
  \centering
  \caption{Built-in experiment presets: baseline, energy window shown on the $E$ axes,
  flux-peak energy used as the default $E$, and measured
  channels~\cite{dune-tdr,nova,t2k,juno,kamland,dayabay}.}
  \label{tab:presets}
  \smallskip
  \begin{tabular}{lccccc}
    \toprule
    Preset & $L$ [km] & $E$ window & $E$ peak & Channels \\
    \midrule
    DUNE & 1300 & 0.5--6~GeV & 2.5~GeV &
      $\nu_\mu\to\nu_e,\ \nu_\mu\to\nu_\mu,\ \nu_\mu\to\nu_\tau$ \\
    NOvA & 810  & 1--3~GeV & 2.0~GeV &
      $\nu_\mu\to\nu_e,\ \nu_\mu\to\nu_\mu$ \\
    T2K & 295 & 0.2--1.5~GeV & 0.6~GeV &
      $\nu_\mu\to\nu_e,\ \nu_\mu\to\nu_\mu$ \\
    JUNO & 52.5 & 1.8--9~MeV & 4~MeV & $\bar\nu_e\to\bar\nu_e$ \\
    KamLAND & 180 & 1.8--9~MeV & 4~MeV & $\bar\nu_e\to\bar\nu_e$ \\
    Daya Bay & 1.66 & 1.8--9~MeV & 4~MeV & $\bar\nu_e\to\bar\nu_e$ \\
    \bottomrule
  \end{tabular}
\end{table}

Beyond the built-ins, users can define, name, and persist their own
experiments from an in-app page: baseline, energy window and
peak (in GeV/km or MeV/m units), measured channels and
$\nu/\bar\nu$ source, and optionally a full set of oscillation-parameter
overrides. Definitions are stored client-side in the browser's
\texttt{localStorage} and registered beside the built-in presets, so
they behave identically to the built-in presets in range scoping, the
reset control, and the preset selector.

%% file: sections/03-views.tex
\section{The 3D views}
\label{sec:views}

Multiple views are necessary in part because neutrino oscillation
involves two bases at once. A neutrino is produced in a pure flavor
state, which is a coherent superposition of mass eigenstates (matter
eigenstates at finite density). Each eigenstate accumulates its own
phase during propagation, and the changing interference between them is
what appears as a changed flavor composition at detection. No single
geometric representation captures both bases: the Statesphere of
section~\ref{sec:sphere} tracks the quantum state in the flavor basis,
while the Phasors view of section~\ref{sec:phasors} decomposes the same
amplitude in the propagation eigenbasis. The other four views present
the measurable probabilities, each along a different dimension of the
parameter space: the Oscillogram (section~\ref{sec:oscillogram}) renders
a probability surface over energy and a second selectable variable, the
Biprobability view (section~\ref{sec:biprob}) shows the
$\nu$--$\bar\nu$ correlation as energy-extruded CP ellipses, the
Flavortube (section~\ref{sec:tube}) follows the full three-flavor
composition along the baseline, and the Worldline view
(section~\ref{sec:loe}) displays the universal $L/E$ dependence and its
breaking by matter effects.

All six views share the same reactive parameter store, the same
oscillation channel (selected in the header: $\nu_\mu\to\nu_e$,
$\nu_\mu\to\nu_\mu$, $\nu_\mu\to\nu_\tau$, or $\nu_e\to\nu_e$), the same
experiment presets, and a common set of interactions. Each view is an
orthographic 3D scene with orbit controls and front/top/side camera
snaps. A per-view ``animate'' control sweeps one chosen variable
(typically $L$, $E$, or $\delta_{\rm CP}$) with play/pause/reset. A hover
probe reports exact numerical values at the cursor, falling back to a
live readout of the animation state when not hovering. One or two
companion 2D panels show the corresponding standard flat plots, redrawn
live and synchronized to the 3D marker; the companion panels are chosen
to present the most common projections for the selected view.

\subsection{Oscillogram}
\label{sec:oscillogram}

\begin{figure}[htbp]
  \centering
  \includegraphics[width=0.49\textwidth]{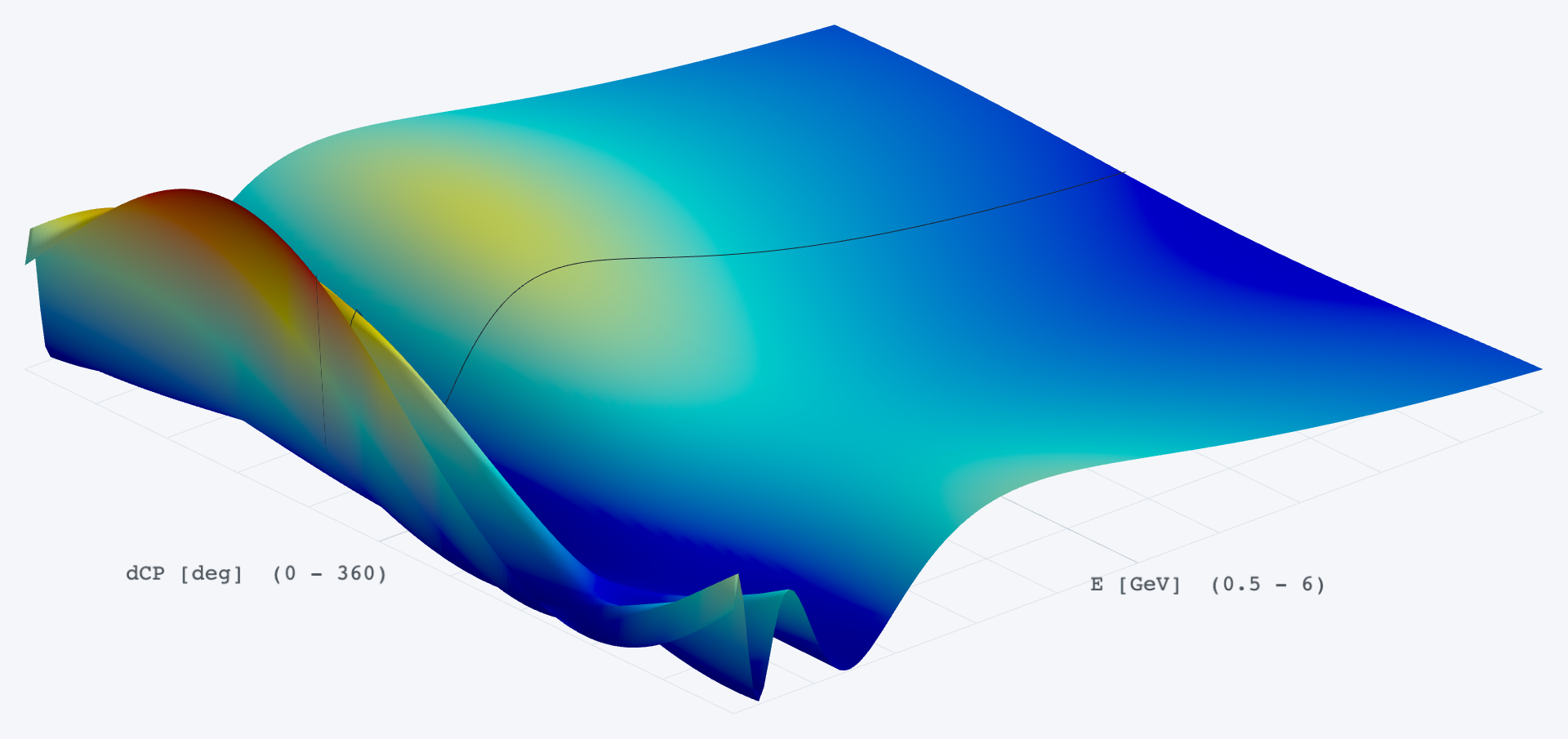}
  \hfill
  \includegraphics[width=0.49\textwidth]{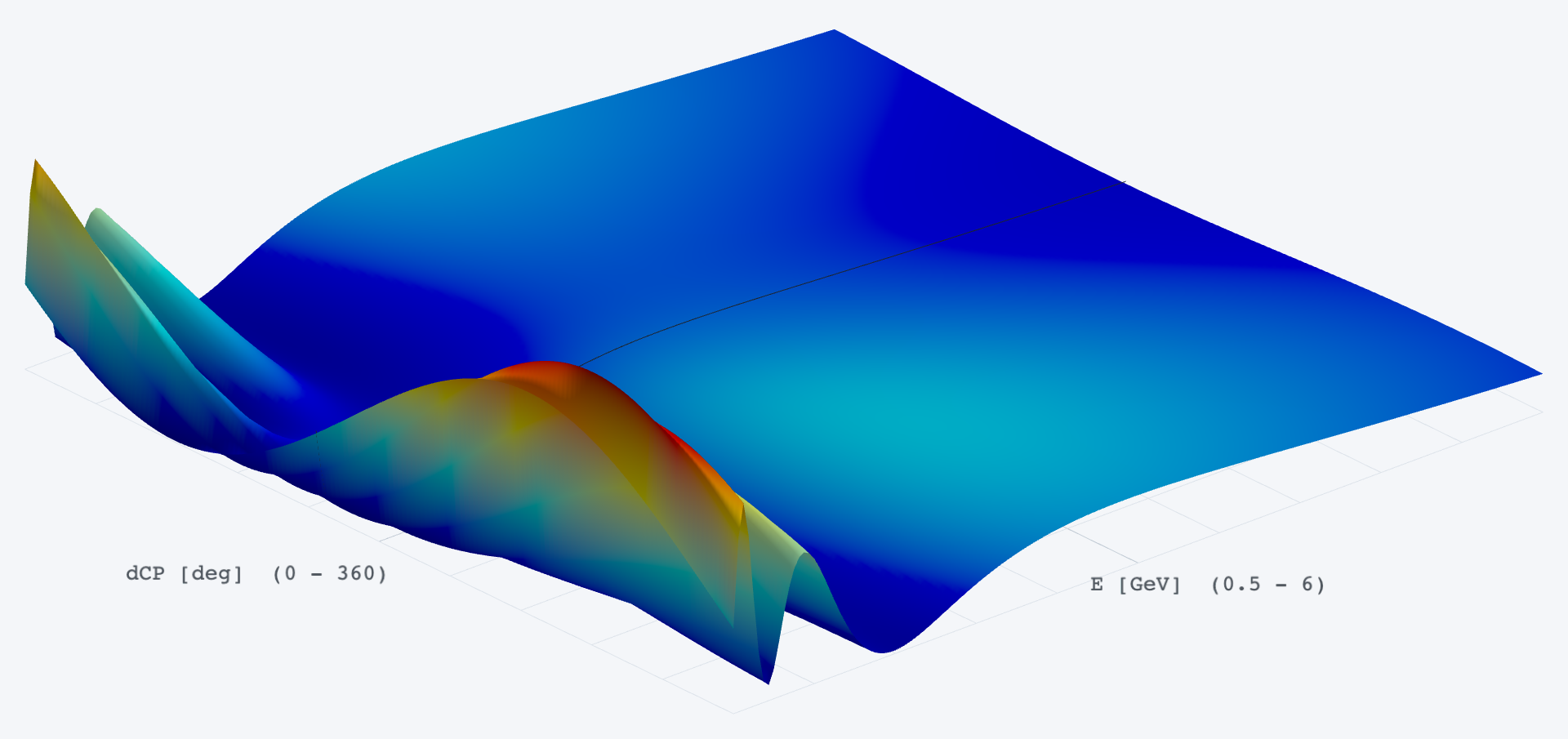}
  \caption{Oscillogram view with the second axis set to $\delta_{\rm CP}$,
  DUNE preset ($L = 1300$~km, $\rho = 2.85$~g/cm$^3$), NuFit~6.1 normal
  ordering, light theme. Left: $P(\nu_\mu\to\nu_e)$; right:
  $P(\bar\nu_\mu\to\bar\nu_e)$, whose ridge sits at the opposite
  $\delta_{\rm CP}$. The thin line on each surface is the marker
  cross-section at $\delta_{\rm CP} = 212^\circ$.}
  \label{fig:oscillogram}
\end{figure}

The Oscillogram view renders the selected channel's probability (default
$P(\nu_\mu\to\nu_e)$) as a height-and-color
surface over the $(E, z)$ plane, where the second axis $z$ is selectable
among $L$ (default, 0--2$\times$baseline), $\delta_{\rm CP}$
(0--360$^\circ$), and $\rho$ (0--5~g/cm$^3$). The default animation sweeps
$\delta_{\rm CP}$, which makes CP violation visible. Companion panels show the $P(E)$ and $P(z)$
line cuts through the marker. The 2D oscillogram commonly seen in the neutrino oscillation literature is recovered as the top-view projection of the surface, rendered with a selectable
color palette (viridis, rainbow, cool--warm, or grayscale). Figure~\ref{fig:oscillogram} shows the $\nu$ and $\bar\nu$
surfaces over the $(E, \delta_{\rm CP})$ plane for the DUNE preset.

\subsection{Biprobability}
\label{sec:biprob}

\begin{figure}[htbp]
  \centering
  \includegraphics[width=0.63\textwidth]{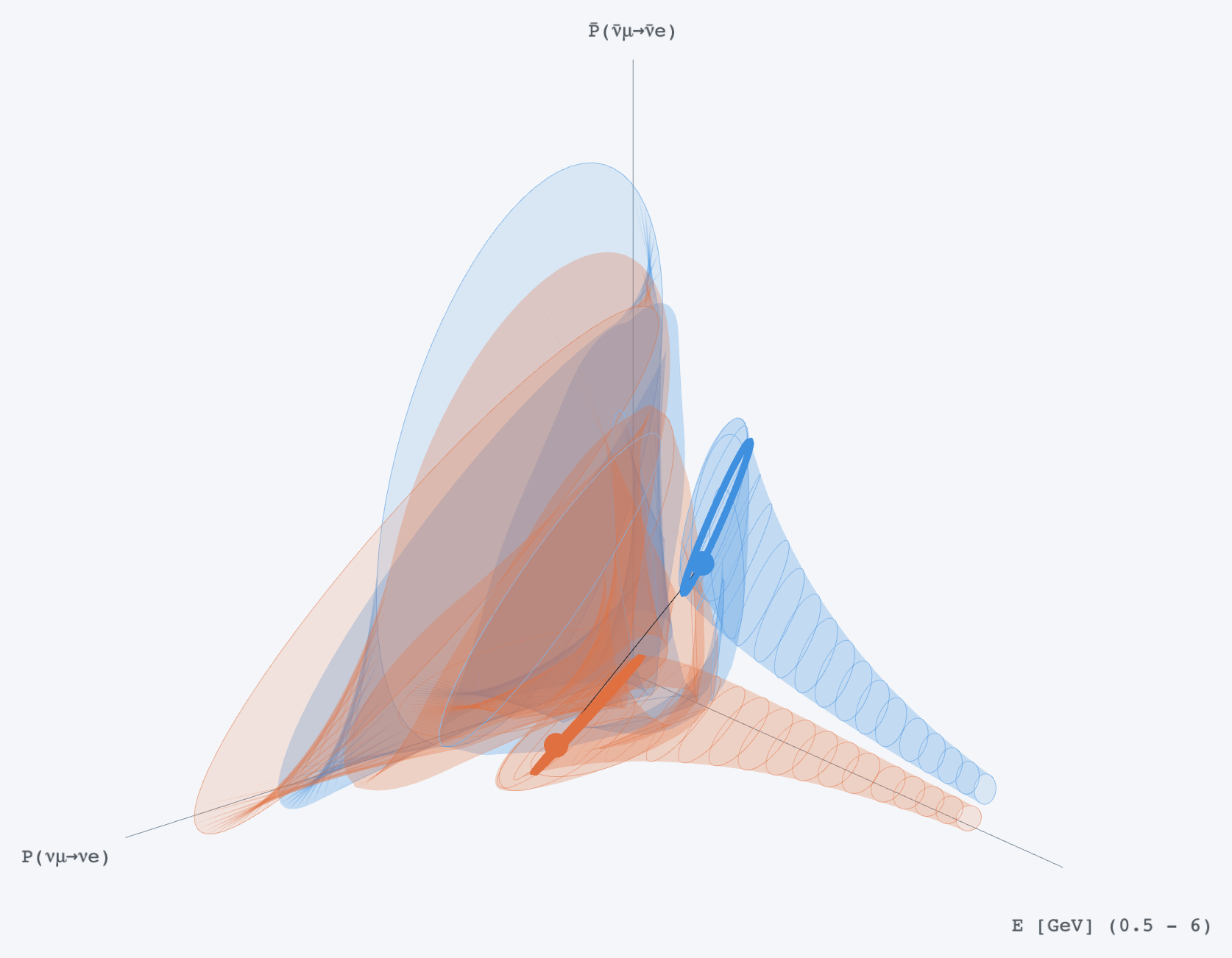}
  \hfill
  \includegraphics[width=0.33\textwidth]{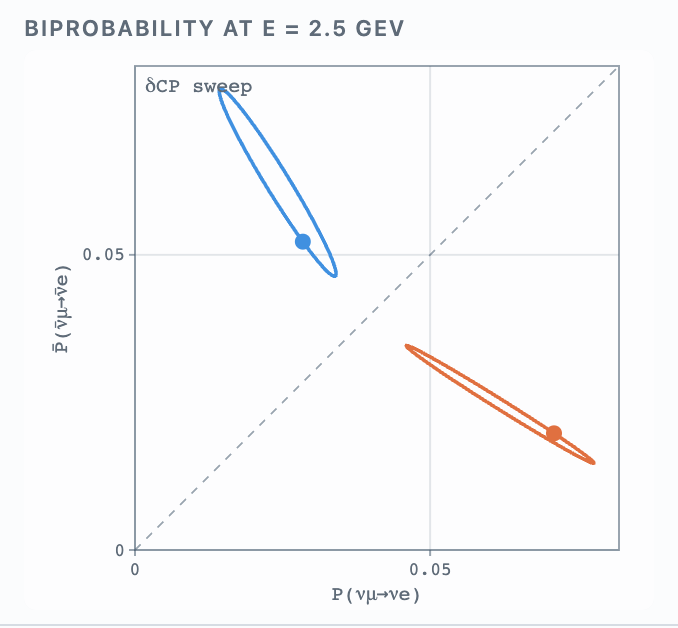}
  \caption{Left: bi-probability tubes for the DUNE preset (NuFit~6.1
  parameters, ``tube surface'' skin enabled), normal ordering in blue and
  inverted in orange, extruded along $E$ from 0.5 to 6~GeV; the
  highlighted rings and dots mark $E = 2.5$~GeV,
  $\delta_{\rm CP} = 212^\circ$. Right: the companion panel at the same
  marker, showing the standard Minakata--Nunokawa CP ellipses, i.e., the
  front-view projection of the tubes at that energy. At 2.5~GeV the two
  ellipses are disjoint: DUNE resolves the ordering at any
  $\delta_{\rm CP}$.}
  \label{fig:biprob}
\end{figure}

The Biprobability view generalizes the Minakata--Nunokawa
diagram~\cite{minakata-nunokawa}: at fixed $E$, the point
(e.g.~$P(\nu_\mu\to\nu_e),\ P(\bar\nu_\mu\to\bar\nu_e)$) traces a closed CP
ellipse as $\delta_{\rm CP}$ runs 0--360$^\circ$. NuGlass extrudes this
family along a third axis, energy, producing one closed tube per mass
ordering (normal and inverted, individually toggleable). The standard
2D diagram is the front-view projection at the marker energy, drawn 1:1 in the companion panel. An optional ``tube surface'' mode skins the
$\delta_{\rm CP} \times E$ grid into a translucent surface. This view gives the mass-ordering degeneracy a geometric interpretation: at energies where the NO
and IO tubes overlap in projection, the ordering cannot
be resolved from that energy alone, and orbiting the camera shows where the tubes separate. Figure~\ref{fig:biprob} shows both tubes for DUNE
with the companion panel's classic 2D diagram at the marker energy.

\subsection{Statesphere}
\label{sec:sphere}

\begin{figure}[htbp]
  \centering
  \includegraphics[width=0.55\textwidth]{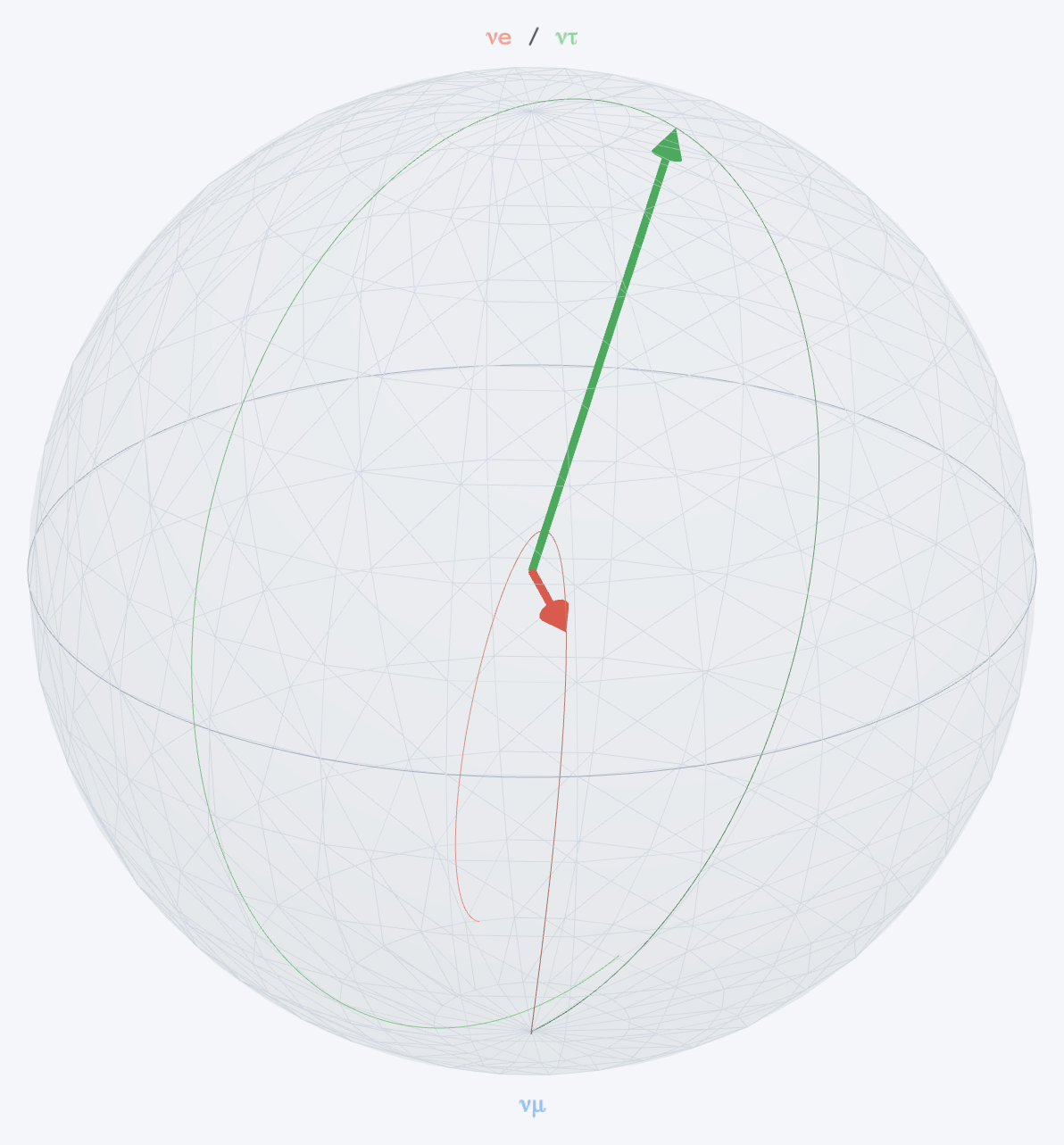}
  \caption{Statesphere in dual-arrow mode, DUNE preset, NuFit~6.1
  parameters, partway through an $L$ sweep ($L = 988$~km). The green
  $\nu_\tau$--$\nu_\mu$ arrow follows a near-great-circle trajectory
  (green curve) from the south pole toward the north pole, staying close
  to the sphere's surface; the red $\nu_e$--$\nu_\mu$ arrow remains short
  and confined near the sphere's center (red trajectory), its length
  $|\vec b\,| = 0.17$ measuring the coherence left in that subspace. The
  exact flavor content at the marker is
  $P_e = 0.049$, $P_\mu = 0.120$, $P_\tau = 0.831$.}
  \label{fig:sphere}
\end{figure}

The Statesphere maps the instantaneous three-flavor state (e.g.~a $\nu_\mu$ at
the source, evolved to distance $L$) onto a Bloch
sphere~\cite{bloch,feynman-two-level} of a chosen two-flavor subspace.
For the $\nu_e$--$\nu_\mu$ subspace the Bloch vector, built from the
flavor amplitudes of eq.~(\ref{eq:amplitude}), is
\begin{equation}
  \vec b = \left(\, 2\,\mathrm{Re}(a_e a^*_\mu),\ 2\,\mathrm{Im}(a_e a^*_\mu),\
  |a_e|^2 - |a_\mu|^2 \,\right),
  \label{eq:bloch}
\end{equation}
with $\nu_e$ at the north pole and $\nu_\mu$ at the south; the
$\nu_\tau$--$\nu_\mu$ subspace is defined analogously, and a dual mode
shows both arrows simultaneously. Equation~(\ref{eq:bloch}) is the
standard two-level construction $b_k = \langle\psi|\sigma_k|\psi\rangle$
written in terms of amplitudes; for a normalized two-flavor state
$\cos(\theta/2)|\nu_\mu\rangle + e^{i\phi}\sin(\theta/2)|\nu_e\rangle$ it
reduces to the familiar polar form
$(\sin\theta\cos\phi,\ \sin\theta\sin\phi,\ \cos\theta)$. The natural
three-flavor generalization replaces the Pauli matrices by the eight
Gell-Mann matrices, $n_a = \langle\psi|\lambda_a|\psi\rangle$;
the state then lives in an 8-dimensional space in which pure states
occupy only a curved 4-dimensional subset rather than a sphere,
so no faithful three-dimensional rendering exists. The Statesphere
therefore projects the exact three-flavor state onto two-flavor
subspaces, at the cost that the projection is not norm-preserving:
$|\vec b| < 1$ measures leakage into the third flavor,
and a pure two-flavor oscillation would trace a circle on the sphere's
surface. The trajectory over the sweep variable is drawn as a
color-graded curve through the sphere's interior;
the geometry it reveals is discussed in section~\ref{sec:sphere-insight},
and figure~\ref{fig:sphere} shows the dual-arrow mode during an $L$
sweep. Companion panels show the three flavor probabilities versus the
sweep variable.

\subsection{Phasors}
\label{sec:phasors}

The Phasors view decomposes the appearance amplitude of
eq.~(\ref{eq:amplitude}) into its three matter-eigenstate terms,
\begin{equation}
  A(L) \equiv a_\beta(L) = \sum_i c_i\, e^{-i\lambda_i L}, \qquad
  c_i = W_{\beta i} W^*_{\mu i},
  \label{eq:phasors}
\end{equation}
where each term is a complex number of fixed magnitude $|c_i|$ rotating
in the complex plane at its own rate $\lambda_i$. Since complex numbers
add like two-dimensional vectors, eq.~(\ref{eq:phasors}) is a vector
sum, and the view draws the three terms head-to-tail as arrows
(phasors) at a marker cross-section, with the resultant $|A|$ as a
separate arrow and the running curves of the partial sums extruded
along the swept variable (e.g.~$L$, $E$, or $\delta_{\rm CP}$). The eigenvalue offset
$\lambda \to \lambda - \lambda_{\rm min}$ is removed (an overall phase
with no effect on $|A|^2$), which pins the lowest-eigenvalue arm as a
stationary reference and makes the \emph{relative} phase velocities
$(\lambda_2 - \lambda_1)$ and $(\lambda_3 - \lambda_1)$ the only visible
rotations. The floor of the scene carries the probability curve
$P = |A|^2$, autoscaled and oriented to match the companion 2D panel
exactly in the top view. A channel selector switches between
$\nu_\mu\to\nu_e$ and $\nu_\mu\to\nu_\tau$; the structural
difference between the two channels is discussed in
section~\ref{sec:phasor-insight}. Unitarity is visible by construction:
at $L = 0$ the three arms form a closed triangle
($\sum_i c_i = \delta_{\beta\mu} = 0$ for $\beta \neq \mu$).

\subsection{Flavortube}
\label{sec:tube}

The Flavortube renders the full flavor composition
(e.g.~$P(\nu_\mu\to\nu_e),\ P(\nu_\mu\to\nu_\mu),\ P(\nu_\mu\to\nu_\tau)$)
along the baseline as a tube whose cross-section is partitioned into
three colored angular sectors of area proportional to the three
probabilities; the constant radius reflects unitarity by construction.
The tube is additionally placed at the shared $E$ on a
depth axis spanning the experiment's beam window, so an energy sweep
translates the tube along a visible axis while its sector pattern
stretches, since the flavors oscillate faster at lower energies.

\begin{figure}[htbp]
  \centering
  \includegraphics[width=0.49\textwidth]{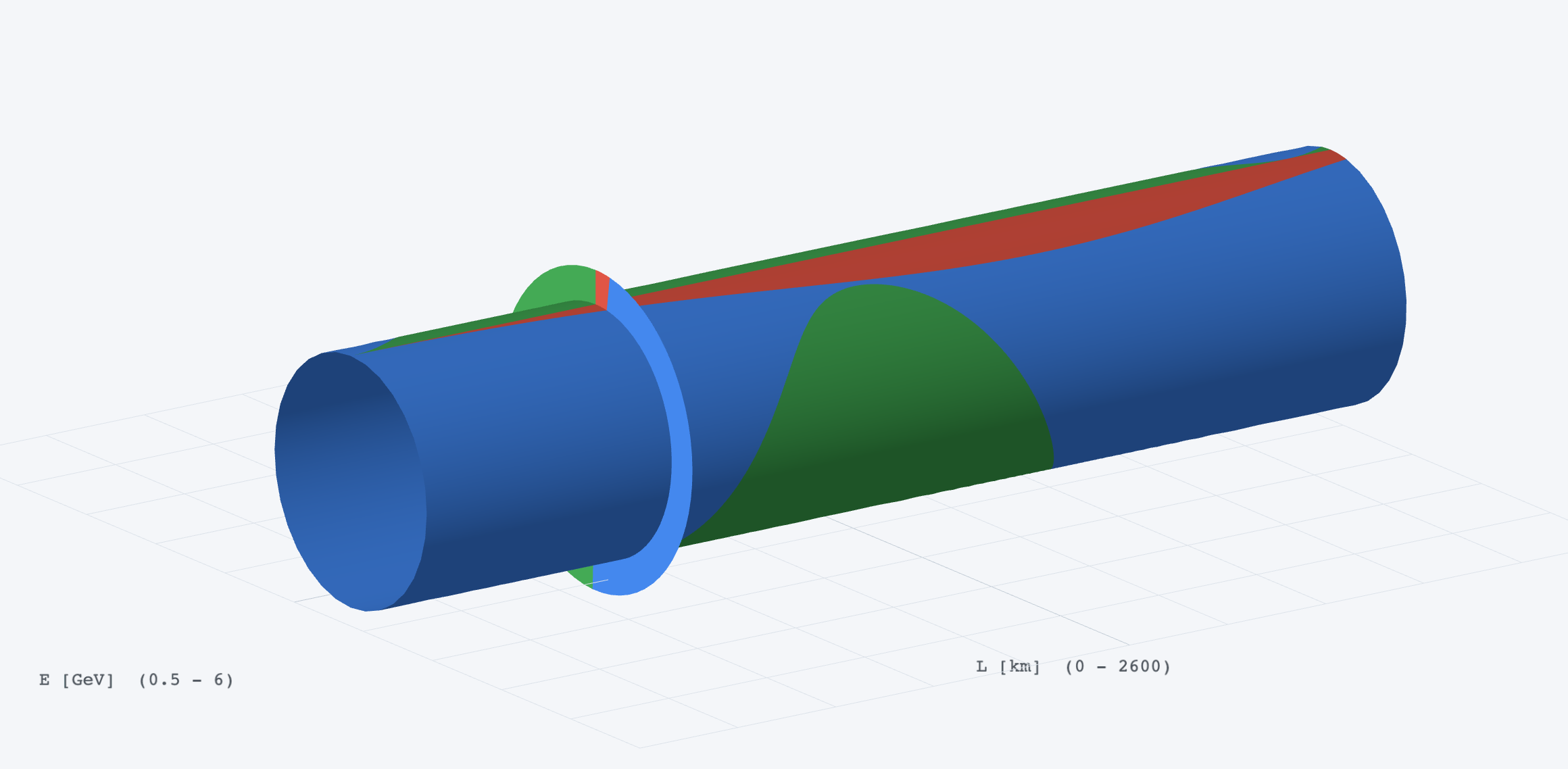}
  \hfill
  \includegraphics[width=0.49\textwidth]{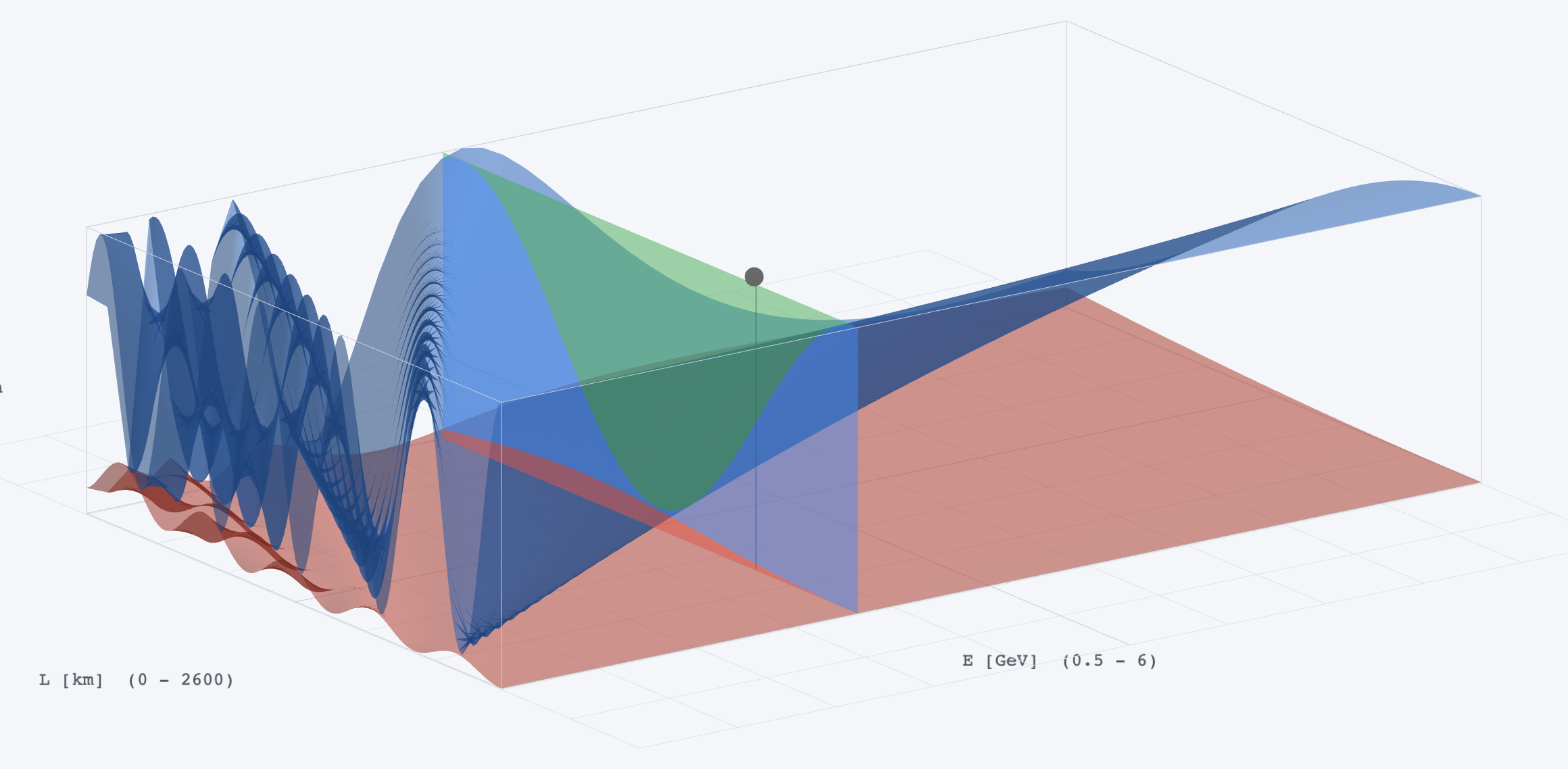}
  \caption{Flavortube view, DUNE preset, NuFit~6.1 parameters, light
  theme. Left: the tube display at $E = 2.5$~GeV with the marker disk at
  $L = 650$~km; the cross-section sectors give the flavor shares (blue
  $\nu_\mu$, red $\nu_e$, green $\nu_\tau$;
  $P_e = 0.023$, $P_\mu = 0.482$, $P_\tau = 0.495$). Right: the
  stacked-bands volume over the same $(L, E)$ ranges, axes matched to the
  oscillogram: the red surface caps the $\nu_e$ share (it is the
  figure~\ref{fig:oscillogram} surface), the blue surface caps
  $\nu_e{+}\nu_\mu$, and the gap up to the box ceiling ($P = 1$) is the
  $\nu_\tau$ share; the translucent slice with the pin marks the shared
  $(E, L)$ point.}
  \label{fig:tube}
\end{figure}

A ``stacked bands'' display transposes the same content into an $(L, E)$
volume with axes matched to the oscillogram ($E$ across, $L$ into
depth): translucent boundary surfaces over the full window (red
capping the $\nu_e$ share, blue capping the cumulative
$\nu_e{+}\nu_\mu$ share) inside a wireframe box whose ceiling
corresponds to total probability 1, so the open gap above the blue
surface is the
$\nu_\tau$ share. A translucent stacked-slice plane at the shared $E$
travels through the volume during an energy animation; the surfaces are
memoized on everything except $E$, so the sweep moves only the slice.
The bottom (red) surface of this volume is identical to the oscillogram
surface of section~\ref{sec:oscillogram}, which makes the
relationship between the two views explicit: the volume is the
unitarity-complete oscillogram, showing where the disappeared flux goes.
Both displays are shown in figure~\ref{fig:tube}.

Two companion panels give the corresponding flat cuts: fractions versus
$L$ at the shared $E$, and the spectrum (fractions versus $E$ at the
marker's current baseline), which updates live as the marker disk moves
along the tube.

The Flavortube is the most outreach-oriented view, showing the beam as a
mixture that evolves between flavors as it travels, while remaining
exactly quantitative. It also complements the Statesphere: the tube
shows the probabilities (the diagonal of the density matrix), while the
sphere shows the coherences of the quantum states.

\subsection{Worldline}
\label{sec:loe}

In vacuum ($W \to U$), the probability $|a_\beta|^2$ of
eq.~(\ref{eq:amplitude}) expands to the standard
oscillation formula
\begin{equation}
  P(\nu_\alpha\to\nu_\beta) = \delta_{\alpha\beta}
  - 4 \sum_{i>j} \mathrm{Re}\!\left(U^*_{\alpha i} U_{\beta i}
      U_{\alpha j} U^*_{\beta j}\right) \sin^2\Delta_{ij}
  + 2 \sum_{i>j} \mathrm{Im}\!\left(U^*_{\alpha i} U_{\beta i}
      U_{\alpha j} U^*_{\beta j}\right) \sin 2\Delta_{ij},
  \label{eq:vacuum}
\end{equation}
which depends on $L$ and $E$ only through the ratio $L/E$ inside the
phases $\Delta_{ij}$ of eq.~(\ref{eq:phase}). Every channel therefore
collapses onto one universal curve in $L/E$. The Worldline view draws these curves
for all four channels against $\log_{10}(L/E)$ over
$30$--$10^5$~km/GeV, so both oscillation frequencies are visible at
once: the first atmospheric extremum near $L/E \approx 500$~km/GeV
($\Delta_{31} = \pi/2$) and the solar valley near
$1.6\times10^4$~km/GeV ($\Delta_{21} = \pi/2$), with the fast
atmospheric wiggle riding on top of it. Dashed markers pin every
built-in experiment at its flux-peak $L/E$: Daya Bay, T2K, NOvA, and DUNE
cluster around the first atmospheric extremum despite baselines spanning
1.66--1300~km, while JUNO sits on the solar valley and KamLAND beyond
it. An optional $\langle\sin^2\Delta_{31}\rangle$ averaging emulates
finite detector resolution, cross-fading the fast wiggle into its mean
as the atmospheric phase grows past $\sim$15--25~rad.

\begin{figure}[htbp]
  \centering
  \includegraphics[width=0.9\textwidth]{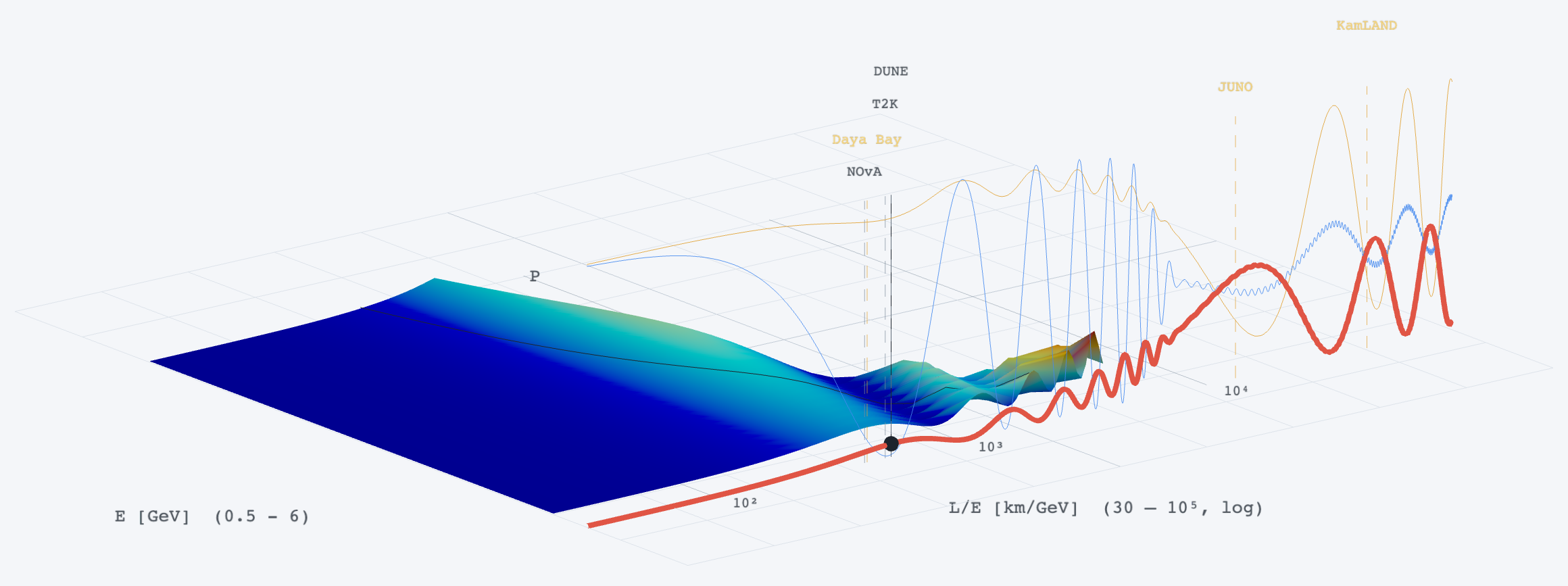}
  \caption{Worldline view, DUNE preset, NuFit~6.1 normal ordering, light
  theme, $E = 2.5$~GeV, $L = 1300$~km (marker dot at
  $L/E = 520$~km/GeV). The front-plane curves are the universal vacuum
  probabilities of eq.~(\ref{eq:vacuum}) against $\log_{10}(L/E)$
  ($\nu_\mu\to\nu_e$ thick, other channels overlaid;
  $\langle\sin^2\Delta_{31}\rangle$ averaging on). The colored
  surface spreads $\nu_\mu\to\nu_e$ across DUNE's energy window,
  computed in matter, and the thin line riding it is the fixed-$L$
  cross-section at $L = 1300$~km.}
  \label{fig:loe}
\end{figure}

The third axis shows how matter effects break this universality: the
matter term of
eq.~(\ref{eq:matterterm}) grows with $E$ at fixed $L/E$, so two
neutrinos with equal $L/E$ but different $E$ no longer oscillate
identically. A surface spreads the selected channel across the
experiment's energy window (each row clipped to
$L \leq 2\times$baseline), computed in matter at the shared $\rho$, with
each energy pulling away
from the vacuum curve by a different amount (figure~\ref{fig:loe}). A
cross-section is drawn on the surface at the experiment's fixed baseline $L$ and coincides with the second
companion panel's spectrum ($P$ versus $E$ at the shared $L$ and
$\rho$, with the vacuum curve dashed); at $\rho = 0$ its front-view
projection lies exactly on the vacuum curve.

%% file: sections/04-architecture.tex
\section{Application architecture}
\label{sec:architecture}

\subsection{Overview}
\label{sec:arch-overview}

\begin{figure}[htbp]
  \centering
  \begin{tikzpicture}[
    font=\small,
    node distance=7mm and 10mm,
    box/.style={draw, rounded corners=2pt, align=center,
                inner sep=5pt, minimum height=8mm},
    store/.style={box, fill=orange!12},
    engine/.style={box, fill=blue!8},
    view/.style={box, fill=green!8},
    shell/.style={box, fill=gray!8},
    arr/.style={-{Stealth[length=2.2mm]}, thick},
  ]
    \node[store] (store) {reactive parameter store\\
      {\footnotesize $E,\ L,\ \rho,\ \delta_{\rm CP},\ \nu/\bar\nu$,
       ordering, $\theta_{ij}$, $\Delta m^2_{ij}$ + per-view state}};
    \node[engine, below left=9mm and 9mm of store] (nufast)
      {NuFast-LBL port\\{\footnotesize probabilities}};
    \node[engine, below right=9mm and 9mm of store] (jacobi)
      {Jacobi engine\\{\footnotesize amplitudes, phasors}};
    \node[view, below=26mm of store] (view)
      {active view module\\
       {\footnotesize \texttt{update()} heavy rebuild, memoized
        $\cdot$ \texttt{tick(dt)} per-frame $\cdot$ \texttt{probe()}}};
    \node[shell, below left=7mm and -24mm of view] (scene)
      {\texttt{SceneBase} 3D stage\\{\footnotesize three.js, camera
       snaps, themes}};
    \node[shell, right=6mm of scene] (panels)
      {companion 2D panels\\{\footnotesize Canvas 2D}};
    \draw[arr] (store) -- node[right=2pt, pos=0.45, font=\footnotesize]
      {reactive reads} (view);
    \draw[arr] (view.190) ++(0,0) -- (scene.north);
    \draw[arr] (view.350) -- (panels.north);
    \draw[arr] (nufast.south) to[bend right=15] (view.165);
    \draw[arr] (jacobi.south) to[bend left=15] (view.15);
    \draw[arr, dashed] (store.west) to[bend right=25]
      node[left=2pt, font=\footnotesize, align=center]
      {UI controls,\\presets, URL} ++(-2.2,-1.2);
  \end{tikzpicture}
  \caption{Application architecture. All UI controls, experiment presets, and URL
  routing mutate the single reactive store; the mounted view module
  reads the store (each read auto-registering as a dependency), calls
  the two engines, and drives the shared 3D stage and its companion 2D
  panels. Heavy geometry rebuilds (\texttt{update()}) and cheap
  per-frame animation (\texttt{tick()}) are separated by contract.}
  \label{fig:arch}
\end{figure}
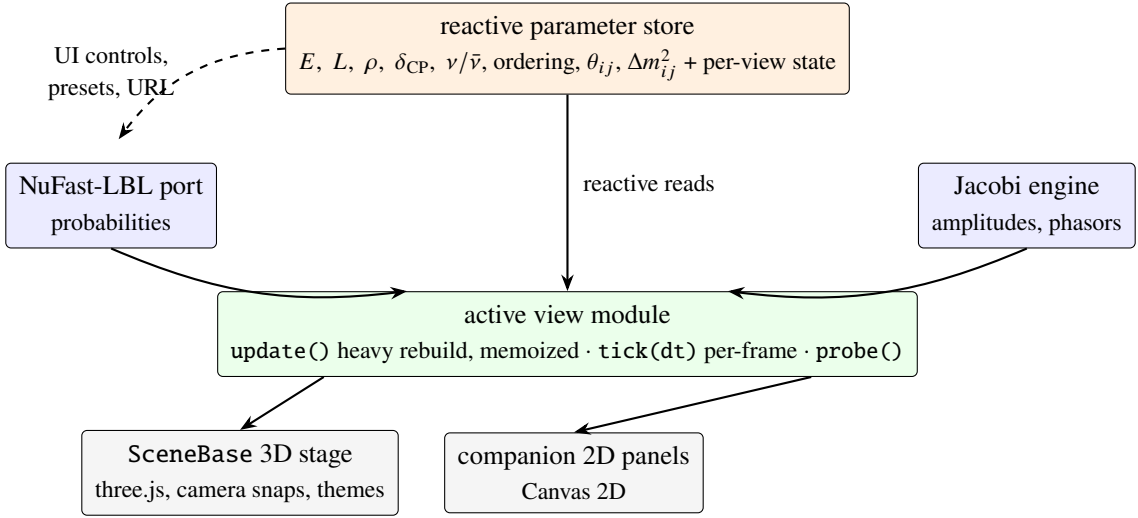

NuGlass is a single-page web application: Vue~3 for reactive
state and UI, three.js~\cite{threejs} (WebGL) for the 3D stages, and
plain Canvas~2D for the companion panels. There is no backend; all
computation runs client-side, so the deployed site is a set of static files
($\approx$190~kB gzipped for the full JavaScript, CSS, and HTML bundle)
served from GitHub Pages. Continuous deployment runs the engine unit
tests and the production build on every push. Figure~\ref{fig:arch}
summarizes the data flow.

The core of the design is a single \emph{reactive parameter store} holding
the shared physics state ($E$, $L$, $\rho$, $\delta_{\rm CP}$,
$\nu/\bar\nu$, ordering, and the six oscillation parameters) plus a small per-view state (selected sweep variable, marker position, toggles). Views
never communicate with each other: they all read the store, so linkage
across views, presets, controls, and URL routing follows from reactivity
rather than from explicit wiring.


Each view is a module exporting a small contract:
\begin{itemize}
\item \texttt{create(container, store)} builds the three.js scene and
  returns \texttt{\{update, tick, probe, dispose\}};
\item \texttt{update()} performs the \emph{heavy} rebuild (surface
  tessellation, trajectory resampling, eigendecompositions over a grid).
  It runs inside a Vue \texttt{watchEffect}: the view re-renders
  when a dependent parameter changes. \texttt{update()} does not
  read the play/marker state, which would otherwise
  retrigger the heavy path every frame;
\item \texttt{tick(dt)} performs the \emph{cheap} per-frame work
  (advancing the animation marker, repositioning arrows and
  cross-sections along precomputed geometry) inside the render loop;
\item \texttt{probe(event)} returns the hover readout string from a
  raycast.
\end{itemize}

This heavy/light split is the central performance decision: a
$\delta_{\rm CP}$ animation on the Oscillogram, for example,
re-tessellates two $96\times96$ surfaces per frame only because
$\delta_{\rm CP}$ is a genuine input to the surface, while an $L$
animation on the Phasors view moves arrows along a cached 1024-sample
decomposition without recomputing any eigensystem. Views additionally
memoize their heavy grids on a serialized key of the parameters that
actually enter them, so toggling a display option never triggers a
physics recomputation. Only the active view is mounted; switching views
disposes the previous scene's GPU resources entirely, so a hidden view
consumes no resources.


A single \texttt{SceneBase} class owns the renderer, an orthographic
camera matched to a reference perspective frustum, orbit controls,
camera-snap tweens to axis-aligned front/top/side projections (per-view
configurable, so that ``top'' always reproduces the companion 2D plot's
orientation; cf.\ section~\ref{sec:phasors}), theme-aware colors for
light/dark modes, billboard text sprites with zoom-invariant on-screen
size and multi-color segments, and the global color palettes. Keeping
every view inside one shell keeps the interaction uniform: a user who
has learned to orbit and read one view can operate all six.


The user interface is a single navigation header (view tabs, play/pause/reset, channel selector, experiment presets, etc.) over a sidebar layout;
the sidebar holds
the shared physics controls, per-view extras, and the companion panels.
Every view has a stable hash-routed URL for direct
linking.

\subsection{Development workflow}
\label{sec:workflow}

The application was developed through an issue-based human--AI
collaboration cycle with a coding agent (Claude Code). Each feature,
fix, or design question is first written up as a GitHub issue. During triage, fully specified issues are routed to autonomous agents, while under-specified or design-heavy tasks stay with human developers for refinement. Larger features go through
design-prototype issues before implementation. The agent implements the
labeled issues and runs the unit tests, and the human reviews the result
both visually in the running app and numerically through the hover
readouts before the issue is closed. The engine unit tests and the
two-engine cross-validation of section~\ref{sec:engines} run in
continuous integration on every push, providing an automated safety net
against possible physics regressions introduced during implementation. We found this
workflow effective for developing scientific software with AI
assistance: it keeps the requirements, the implementation history, and
the review discussion in one public place, and it confines the agent's
autonomy to tasks whose acceptance criteria are stated in advance.

\subsection{Performance}
\label{sec:performance}

The application requirements are modest: any browser with WebGL~2.0 and modern
JavaScript (ES2020) support, which includes all mainstream desktop and
mobile browsers released since about 2021, runs the app without
plug-ins or installation. The app has been tested on modern desktop and
mobile browsers, including phones and tablets. It is also packaged as a
progressive web app: it can be installed from the browser as a
standalone application, and a service worker caches the full static
bundle on the first visit, after which the app runs without a network
connection, e.g.\ for classroom or outreach use.

On an Apple M3~Max laptop (Chromium, production build), the app
cold-loads in well under a second. Views whose animation follows the
cheap path (repositioning markers and arrows along cached geometry)
sustain 45--60 frames per second. Views whose
animated variable is a genuine input to the surface run the heavy path
each step and are correspondingly slower: the oscillogram's default
$\delta_{\rm CP}$ sweep re-tessellates two $96\times96$ surfaces per
frame (an eigendecomposition per vertex) at roughly 20~fps, and the
Flavortube's $L$ sweep, which rebuilds the tube's sector geometry, runs
at roughly 12~fps. This is a deliberate trade-off that keeps the engines
exact while remaining fluid enough for live
demonstration. Static interaction (orbiting, hovering, slider drags on a paused scene) always runs at the display refresh rate.

%% file: sections/05-insights.tex
\section{Example physics observations}
\label{sec:insights}

While the probability views (Oscillogram, Biprobability, Flavortube,
Worldline) are more familiar to researchers, this section showcases two
observations from the quantum-state views in detail. Neither is new
physics (both follow from the textbook three-flavor formalism), but in
each case the 3D representation makes directly visible a structural
feature that is hard to read off standard 2D projections. They provide
concrete evidence that the added dimension carries physics information,
not just decoration.

\subsection{Statesphere: flavor evolution is a squashed, near-planar
precession}
\label{sec:sphere-insight}

In an exact two-flavor system the Bloch vector obeys the
Feynman--Vernon--Hellwarth precession
equation~\cite{feynman-two-level}: it rotates rigidly about the fixed
axis defined by the Hamiltonian,
$\hat n = (\sin 2\theta_{\rm eff},\ 0,\ \cos 2\theta_{\rm eff})$, at the
oscillation frequency. Three-flavor evolution projected onto a two-flavor
subspace is not exactly a precession, but the app shows how much of that
geometry survives in the three-flavor case and where it breaks down.

\paragraph{The $\nu_\tau$--$\nu_\mu$ arrow traces a near-great circle
between the poles.} With $\theta_{23} \approx 45^\circ$, the effective
Hamiltonian axis in the $\nu_\tau$--$\nu_\mu$ subspace lies nearly in the
equator, so a state born at the south pole ($\nu_\mu$) precesses on a
near-maximal cone: a circle passing close to both poles, confined to a
single vertical plane (the azimuth, the relative phase between
$a_\tau$ and $a_\mu$, stays locked while the polar angle sweeps).
The circle is not exact: because $\theta_{13} \neq 0$, a small
probability always leaks into $\nu_e$, so $|\vec b| < 1$ and the arrow
never reaches the north pole (a pure $\nu_\tau$ state) even for
$\theta_{23} = 45^\circ$ exactly; the trajectory runs slightly inside
the sphere's surface.
The green arrow swinging pole-to-pole within a single plane is
nonetheless a direct display of near-maximal atmospheric mixing, and the
two deviations are geometrically distinct: any octant deviation of
$\theta_{23}$ from $45^\circ$ tilts the precession axis, making the
circle miss the north pole by a fixed margin along the surface, while
the $\theta_{13}$-induced leakage pulls the trajectory radially inward.

\paragraph{The $\nu_e$--$\nu_\mu$ arrow lives in the bottom half of the
sphere.} Its polar coordinate is $b_z = P_e - P_\mu$. Because
$\theta_{13}$ is small, the appearance probability $P_e$ never exceeds
$\sim$0.1 at long-baseline energies, while $P_\mu$ starts at 1: the red
arrow starts at the south pole and, over an $L$ sweep, climbs only
marginally while shrinking toward the center of the sphere. The climb
is bounded by $\theta_{13}$, while the shrinkage measures the leakage
into the third flavor.
Since the projection of the 8-dimensional three-flavor state onto a
two-flavor subspace is not norm-preserving, $|\vec b| < 1$ exactly
measures the coherent leakage into $\nu_\tau$. The dual-arrow mode makes the
complementarity explicit: as the red ($\nu_e$--$\nu_\mu$) arrow shrinks,
the green ($\nu_\tau$--$\nu_\mu$) arrow sweeps through its great circle,
and the length missing from one arrow appears in the motion of the
other.

\subsection{Phasors: matter collapses $\nu_\mu\to\nu_\tau$ into a
two-phasor interference}
\label{sec:phasor-insight}

The appearance amplitude in matter is $A = \sum_i c_i e^{-i\lambda_i L}$
with $c_i = W_{\beta i} W^*_{\mu i}$ (eq.~(\ref{eq:phasors}));
unitarity closes the triangle
($\sum_i c_i = 0$ for $\beta \neq \mu$), which the view shows as the
three arms forming a closed loop at $L = 0$. Each arm's length is the
product of two flavor contents of matter eigenstate $i$: how much
$\nu_\beta$ it contains times how much $\nu_\mu$.

\paragraph{$\nu_\mu\to\nu_e$ is a three-phasor system with full
$\delta_{\rm CP}$ sensitivity.} All three arms are of comparable
magnitude, the
triangle is non-degenerate, and changing $\delta_{\rm CP}$ visibly
rotates the arms' relative orientations, deforming $|A|^2$. This is the
phasor picture of the standard statement that CP violation is an
interference effect between the solar and atmospheric frequencies.

\begin{figure}[htbp]
  \centering
  \includegraphics[width=0.49\textwidth]{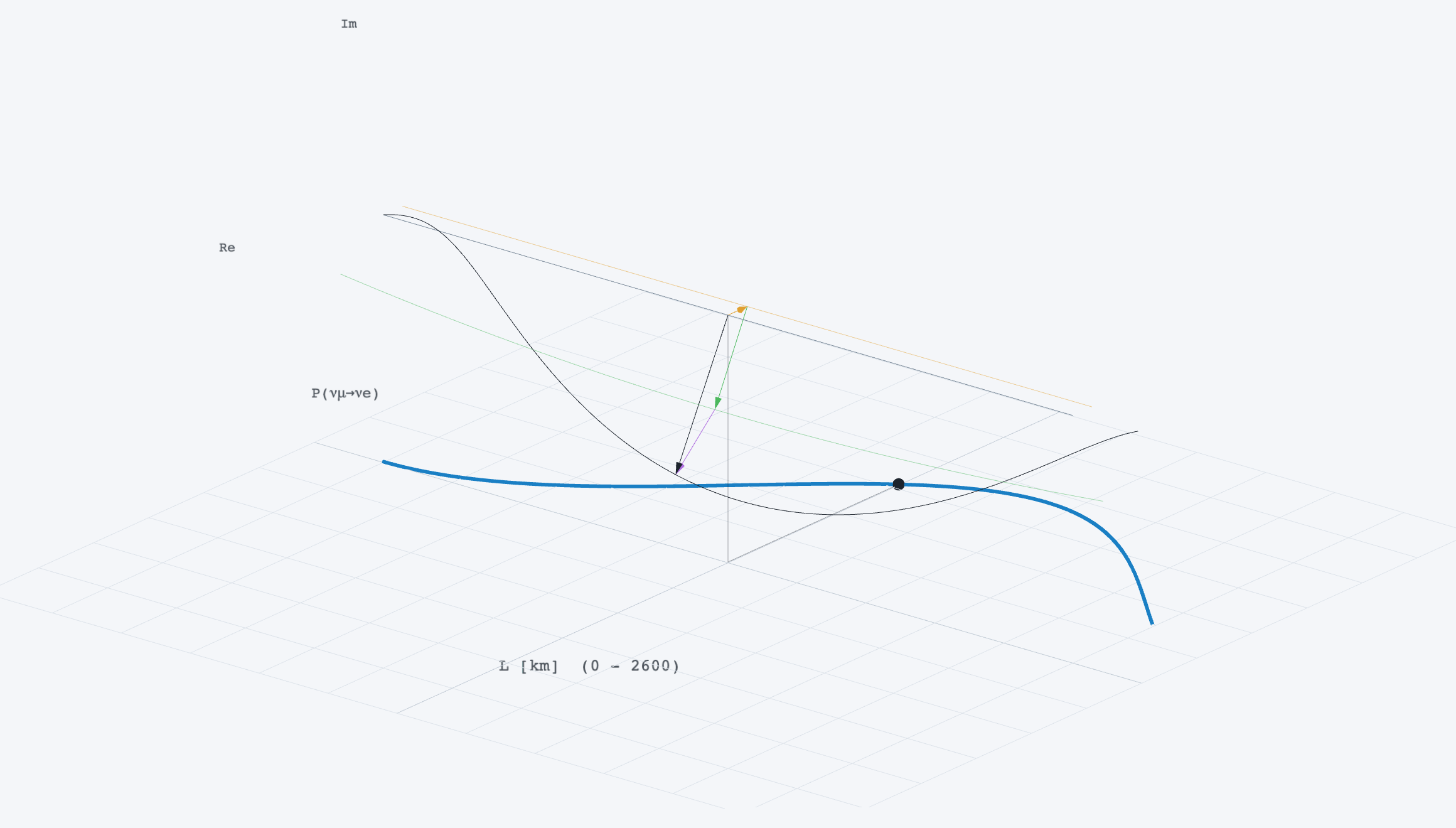}
  \hfill
  \includegraphics[width=0.49\textwidth]{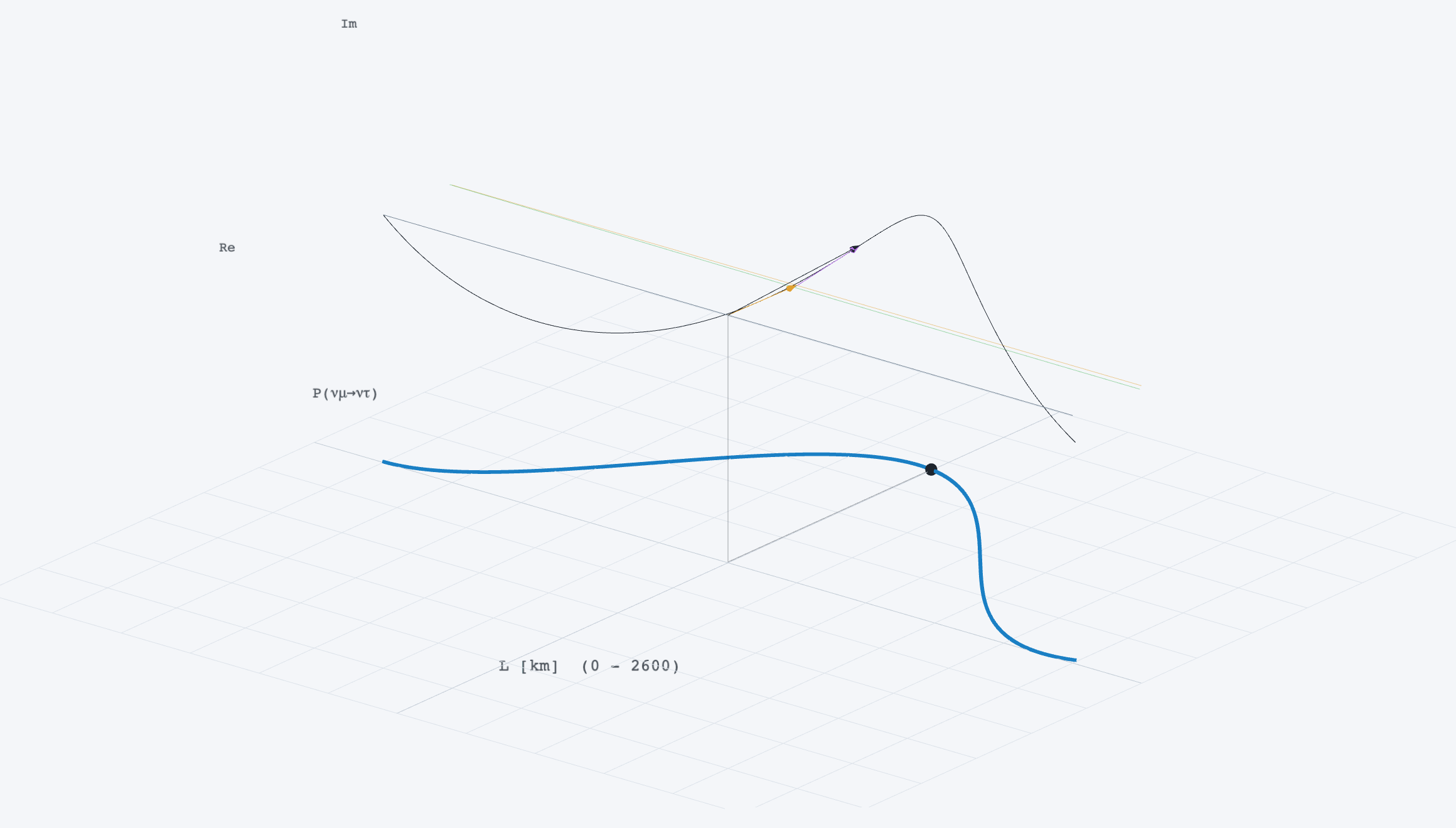}
  \caption{Phasor view at the same parameter point ($E = 2.5$~GeV,
  DUNE preset, NuFit~6.1 normal ordering, marker at $L = 1300$~km).
  Left: $\nu_\mu\to\nu_e$, a genuine three-phasor system
  ($|c_i| = 0.049,\ 0.171,\ 0.128$); the three arms are drawn
  head-to-tail from the origin with the black resultant $|A|$, the black
  curve tracing $|A|(L)$, and the floor carrying $P = |A|^2$ (blue,
  autoscaled). Right: $\nu_\mu\to\nu_\tau$ at the identical point,
  where the middle arm has collapsed ($|c_i| = 0.497,\ 0.018,\ 0.481$)
  and the amplitude is a two-phasor interference.}
  \label{fig:phasors}
\end{figure}

\paragraph{$\nu_\mu\to\nu_\tau$ degenerates to two phasors in matter.} At DUNE-like parameters ($E = 2.5$~GeV, $\rho = 2.85$~g/cm$^3$,
NuFit~6.1 best fit) the oscillation amplitude reads $|c_1| = 0.497$, $|c_2| = 0.018$, $|c_3| = 0.481$ (figure~\ref{fig:phasors} and
table~\ref{tab:phasors}): the middle arm has collapsed by a
factor of $\sim$25 relative to its neighbors. The reason is a matter
effect with a clean geometric signature: at these energies (well above
the solar level crossing, below the atmospheric one) the intermediate
matter eigenstate is almost purely $\nu_e$. Its $\nu_\mu$ content and its
$\nu_\tau$ content are therefore \emph{both} near zero, and $c_2$,
their product, is doubly suppressed. In vacuum the same quantity is
$|U_{\tau 2}||U_{\mu 2}| = 0.34$; sweeping the density control
toward zero shows the middle arm regrowing continuously. The $\nu_\mu$
and $\nu_\tau$ flavors are shared almost entirely between the outer two
matter eigenstates, so the channel reduces to
\begin{equation}
  P(\nu_\mu\to\nu_\tau) \approx 4\,|c_1||c_3|
  \sin^2\!\left(\frac{(\lambda_3 - \lambda_1)L}{2}\right).
  \label{eq:twophasor}
\end{equation}
This is a clean two-frequency interference peaking near
$\sin^2 2\theta_{23}$; with no third arm of the triangle left to rotate,
it has almost no $\delta_{\rm CP}$ sensitivity.

\begin{table}[htbp]
  \centering
  \caption{Phasor arm magnitudes $|c_i| = |W_{\beta i} W^*_{\mu i}|$ at
  the figure~\ref{fig:phasors} parameter point ($E = 2.5$~GeV, NuFit~6.1
  normal ordering), in DUNE matter ($\rho = 2.85$~g/cm$^3$) and in
  vacuum. Matter suppresses $|c_2|$ of the $\nu_\mu\to\nu_\tau$ channel
  by an order of magnitude while leaving the outer arms nearly equal;
  the $\nu_\mu\to\nu_e$ channel keeps three comparable arms.}
  \label{tab:phasors}
  \smallskip
  \begin{tabular}{lcccccc}
    \toprule
    & \multicolumn{3}{c}{$\nu_\mu\to\nu_e$}
    & \multicolumn{3}{c}{$\nu_\mu\to\nu_\tau$} \\
    \cmidrule(lr){2-4}\cmidrule(lr){5-7}
    & $|c_1|$ & $|c_2|$ & $|c_3|$ & $|c_1|$ & $|c_2|$ & $|c_3|$ \\
    \midrule
    matter, $\rho = 2.85$~g/cm$^3$ & 0.049 & 0.171 & 0.128
                                   & 0.497 & 0.018 & 0.481 \\
    vacuum                          & 0.276 & 0.360 & 0.102
                                   & 0.154 & 0.340 & 0.488 \\
    \bottomrule
  \end{tabular}
\end{table}

%% file: sections/06-discussion.tex
\section{Discussion}
\label{sec:discussion}

The app has three overlapping and complementary roles. As a
\emph{research companion}, it provides quick, quantitative answers
(``where do the NO/IO bi-probability tubes overlap for NOvA?'', ``how
large is $|c_2|$ at 4~GeV?'') with exact engines and current global-fit
defaults; the hover probes and companion 2D panels are designed so
that qualitative impression can be read back as a number. As a
\emph{teaching instrument}, the linked views take a student through
successive levels of abstraction, from probabilities (Flavortube,
Oscillogram, Worldline) to correlations (Biprobability), coherences
(Statesphere), and amplitudes (Phasors), all driven by the same
parameter store so that the connections between representations are
shown directly. As an \emph{outreach exhibit}, the
animation-first defaults, experiment presets, and zero-installation deployment make it suitable for showcasing the science to the general public.

There are several limitations of the current app:
\begin{itemize}
\item \textbf{Constant density.} Both engines assume constant matter
  density; there is no PREM profile, so atmospheric/solar trajectories
  and very long baselines are out of scope. This is the standard
  approximation for the accelerator baselines that the app targets.
\item \textbf{Inverted ordering as a sign flip.} IO is approximated by
  flipping the sign of $\Delta m^2_{31}$ at the NO best-fit magnitudes;
  the NuFit IO column differs slightly in the other parameters.
  This is acceptable for visualization but should be kept in mind for quantitative use.
\item \textbf{Initial flavor follows the channel.} The initial flavor is
  $\nu_\mu$ for the beam channels and $\nu_e$ for reactor disappearance;
  $\nu_\tau$ sources and atmospheric flavor mixes are not represented.
\item \textbf{No experimental realism.} Fluxes, cross sections,
  resolutions, and statistical uncertainties are intentionally absent:
  the app visualizes only the oscillation probabilities.
\item \textbf{Numerical, not symbolic.} Statements such as the
  eigenstate flavor content in section~\ref{sec:phasor-insight} are read
  off the numerical eigenvectors; the app does not expose analytic
  matter-perturbation formulas.
\end{itemize}


All of the above limitations point to natural future extensions, and contributions from the community are encouraged given the open-source nature of the project.

%% file: sections/07-conclusions.tex
\section{Conclusions}
\label{sec:conclusions}

We have presented NuGlass, an open-source, browser-native application
that renders three-flavor neutrino oscillations in matter as six linked,
interactive 3D views (oscillogram surface, energy-extruded
bi-probability tubes, Bloch-sphere state projection, matter-eigenstate
phasor decomposition, flavor-composition tube, and $L/E$ worldline with
matter-effect degeneracy breaking), driven by two
independently implemented engines (a NuFast-LBL port and an exact
complex-Jacobi amplitude solver) cross-validated to the $10^{-7}$ level
in probability. The architecture is a small, static, fully
client-side bundle with a per-view contract that cleanly separates
physics rebuilds from animation, making the tool easy to deploy and
extend.

Beyond serving as an interactive calculator, the added dimension
carries real physics content: the near-planar pole-to-pole precession of
the flavor state displays near-maximal $\theta_{23}$ geometrically, and the
collapse of the $\nu_\mu\to\nu_\tau$ amplitude triangle into a two-phasor
interference makes a specific matter effect directly visible and continuously tunable. We offer these as evidence
that interactive 3D representation is an underused channel for
communicating oscillation phenomenology, to experts and to the public
alike.

The application is deployed at \url{https://czczc.github.io/nuglass/}
with source, issue tracker, and citation metadata at
\url{https://github.com/czczc/nuglass}.